\documentclass[journal=aamick,manuscript=letter,layout=twocolumn]{achemso}

\usepackage[version=3]{mhchem} 
\usepackage{graphicx}
\usepackage{dcolumn}
\usepackage{bm}
\usepackage[mathlines]{lineno}
\usepackage{amssymb}
\usepackage{amsmath}
\usepackage{soul,color}
\usepackage[colorlinks,linkcolor=blue,anchorcolor=blue,citecolor=blue,urlcolor=blue]{hyperref}
\usepackage{dcolumn}

\usepackage{threeparttable}  
\usepackage{cmll}
\usepackage{multirow}
\usepackage{booktabs}
\usepackage{color}

\title{Ultralow lattice thermal conductivity of monolayer penta-silicene and penta-germanene}

\author{Zhibin Gao}
\email{zhibin.gao@nus.edu.sg}
\affiliation{Department of Physics, National University of Singapore,
             Singapore 117551, Republic of Singapore}
\author{Zhaofu Zhang}
\affiliation{Department of Engineering, Cambridge University, Cambridge,
             CB2 1PZ, United Kingdom}
\author{Gang Liu}
\email{liugang8105@gmail.com}
\affiliation{School of Physics and Engineering, Henan University of
             Science and Technology, Luoyang 471023, China}
\author{Jian-Sheng Wang}
\affiliation{Department of Physics, National University of Singapore,
             Singapore 117551, Republic of Singapore}

\date{\today}

\begin{document}

\begin{abstract}
We study the lattice thermal conductivity of two-dimensional (2D)
pentagonal systems, such as penta-silicene and penta-germanene.
Penta-silicene has been recently reported\cite{guo2019lattice},
while the stable penta-germanene, belonging to the same group IV
element, is revealed firstly by our \textit{ab initio} calculations.
We find that both penta-silicene and penta-germanene at room temperature
have ultralow lattice thermal conductivities $\kappa$ of 1.29~W/mK and
0.30~W/mK respectively . To the best of our knowledge, penta-germanene may
have the lowest $\kappa$ in 2D crystal materials. We attribute
ultralow $\kappa$ to the weak phonon harmonic interaction and strong
anharmonic scattering. A small phonon group velocity, a small Debye
frequency, a large Gr\"{u}neisen parameter, and a large number of modes
available for phonon-phonon interplay together lead to the ultralow
$\kappa$ of penta-silicene and penta-germanene. These discoveries
provide new insight into the manipulation of ultralow $\kappa$ in
2D materials and highlight the potential applications of designing
silicon and germanium based high thermoelectric materials\cite{hu2019silicon}.
\end{abstract}

\section{Introduction}
%
Except for oxygen, silicon is the most abundant element in Earth's
crust (27.7\%). Since the 1950s, the development of computer technology
and the microelectronics industry is based on silicon chips. According
to Moore's law, the number of transistors consisting of a chip will
be doubled around every 18 months. However, this rule may become
deviated due to the tremendous difficulty of scaling down the
conventional silicon transistors\cite{mathur2002nanotechnology}.
Recently, nanomaterials, similar to thin-films, have attracted much
attention for more superior transistors\cite{franklin2015nanomaterials}.

Silicene, as a representative 2D materials of group IV elements, has been
successfully made into field-effect transistors\cite{le20152d}. However, the
main drawback for silicene as a transistor material is the absence of a band
gap for the on/off ratio\cite{cahangirov2009two}. In 2014,
penta-graphene with a Cairo pentagonal tiling, as a new carbon
allotrope, was firstly predicted and has focused attention on the pentagonal
system\cite{zhang2015penta,wu2016hydrogenation,liu2016disparate,oyedele2017pdse2}.
However, penta-silicene, as a new successor of penta-graphene, has
been reported to be unstable due to imaginary frequencies of phonon
dispersion\cite{ding2015hydrogen,aierken2016first}.
Recently, \textit{Guo et al.}\cite{guo2019lattice} proposed that
tilting the Si dimers would reduce the Coulomb interaction and stabilize it.
Besides, they also reported the strong ferroelectricity with a high Curie
temperature of 1190~K of penta-silicene.
However, a study of the lattice conductivity is lacking.
Moreover, it is also interesting to find a stable penta-germanene in group IV element.

In this study, we explore the lattice thermal conductivities and thermal
transport properties of penta-silicene and penta-germanene based on
\textit{ab initio} calculations. We find that both penta-silicene and
penta-germanene have ultralow $\kappa$
of 1.29~W/mK and 0.30~W/mK at 300~K, which are much lower than
the penta-graphene of 645~K. These may be the lowest $\kappa$ in 2D
crystal materials based on our collected data. We attribute both
ultralow $\kappa$ to the weak phonon harmonic interaction and strong
anharmonic scattering. We hope these discoveries will make a contribution
to the thermal transport of 2D materials and silicon's energy harvesting
power\cite{hu2019silicon}.

\section{Results and discussion}
\begin{figure}[t!]
\includegraphics[width=1.0\columnwidth]{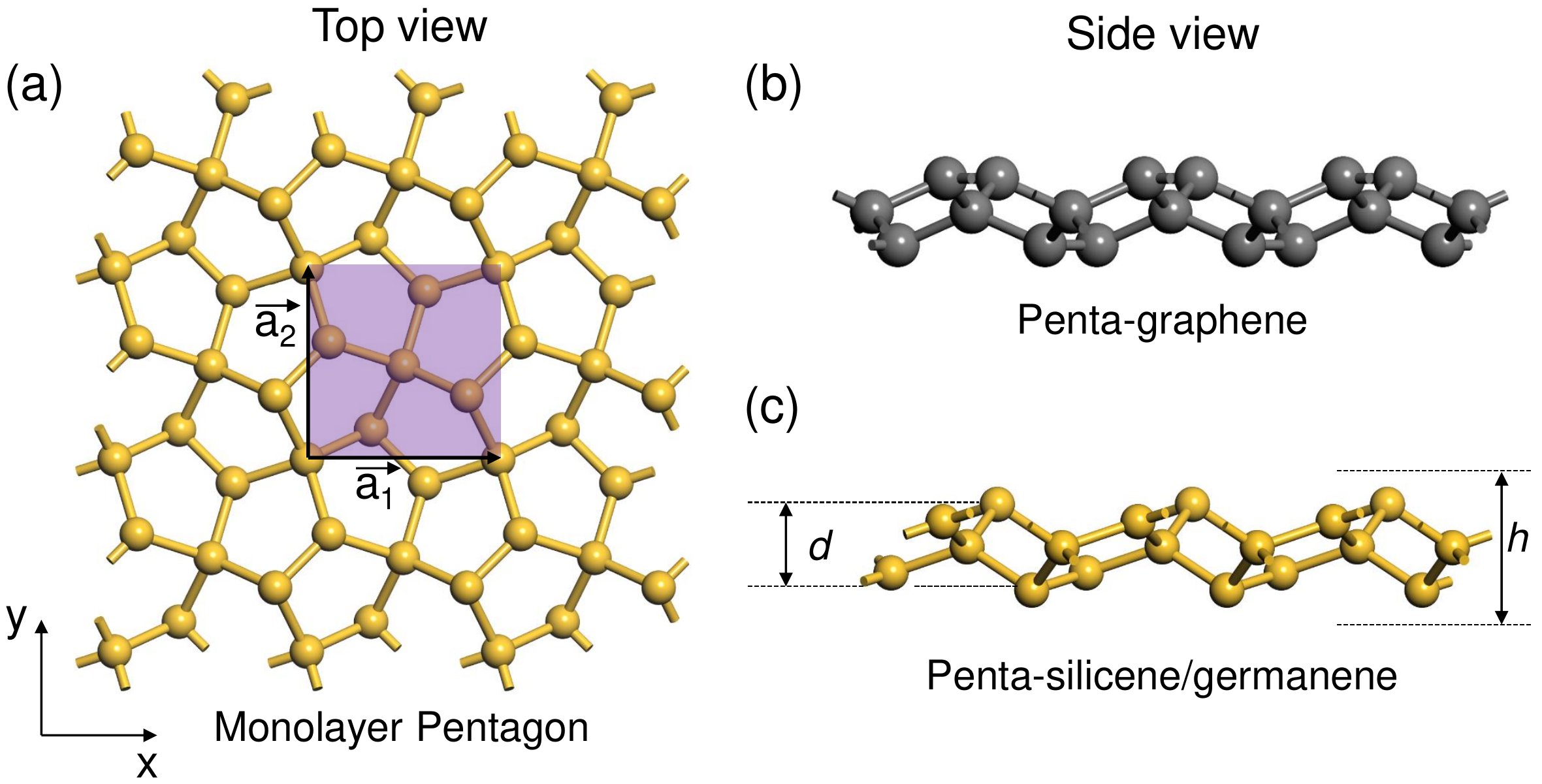}
\caption{
Ball-and-stick model of the 2D pentagonal
system of a 3 $\times$ 3 supercell in (a) top
view (penta-graphene, penta-silicene and penta-germanene).
Side views of penta-graphene are shown in (b) and of
penta-silicene or penta-germanene are shown in (c).
The primitive cell is indicated by the purple shading in (a).
$\vec{a_1}$ and $\vec{a_2}$ are the corresponding lattice vectors
and \textit{d} is the buckling distance. The effective thickness
\textit{h} is defined as the summation of \textit{d} and two van
der Waals radii of the outmost surface atom of
structure\cite{gao2017novel}.
}\label{fig1}
\end{figure}

\begin{figure*}[t!]
\centering
\includegraphics[width=1.8\columnwidth]{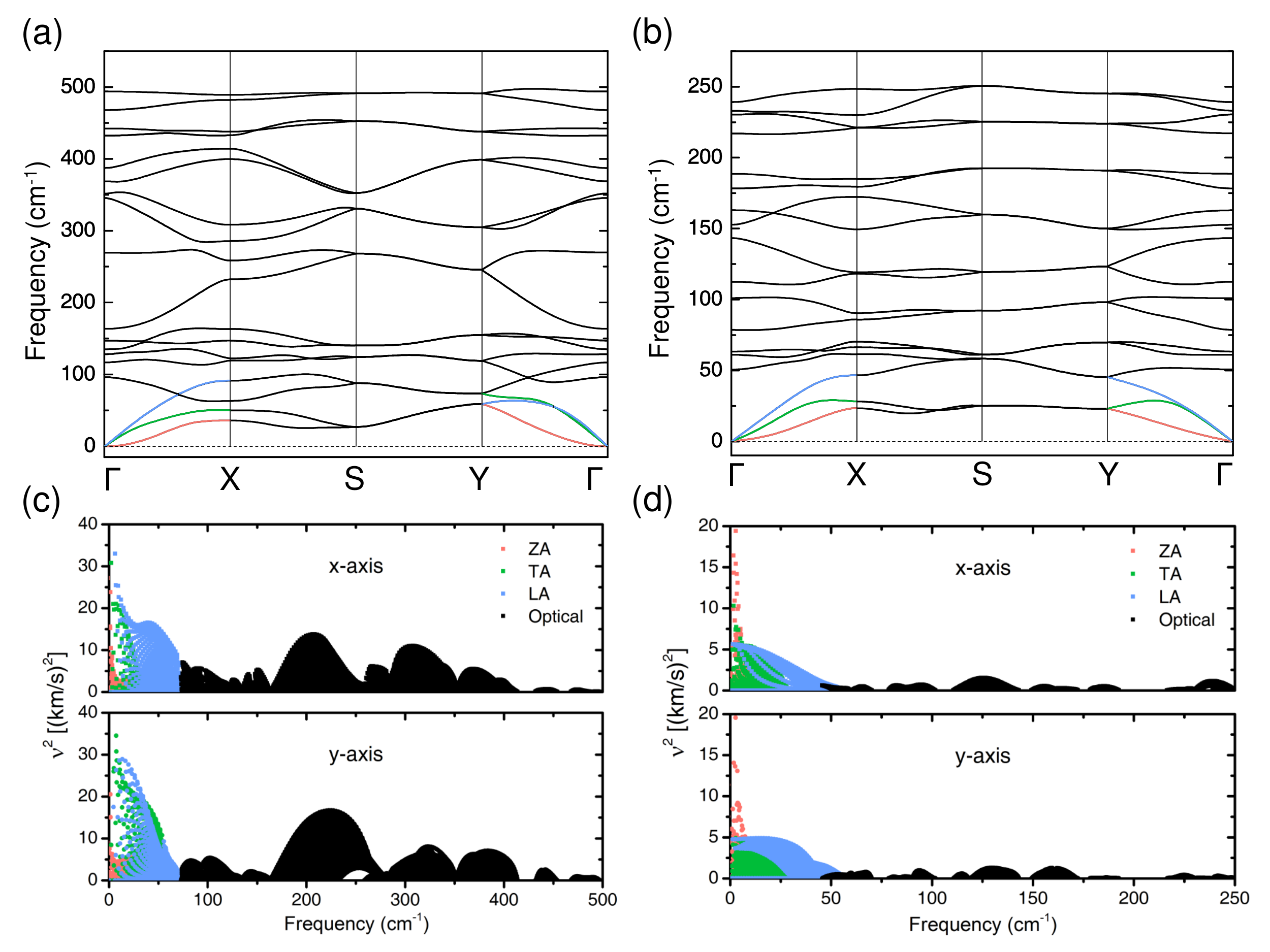}
\caption{Phonon dispersions of (a) penta-silicene and (b)
penta-germanene. In the first Brillouin zone, the high symmetry
$k$ points are: $\Gamma$(0 0 0), X(1/2 0 0), S(1/2 1/2 0)
and Y(0 1/2 0). Three acoustic phonon branches, which correspond
to an out-of-plane (ZA) mode, an in-plane transverse (TA) mode,
and an in-plane longitudinal (LA) mode, are marked. Phonon velocity
squared of (c) penta-silicene and (d) penta-germanene along x-axis
and y-axis with mode resolution.
}\label{fig2}
\end{figure*}

The optimized structure of the 2D pentagonal system (penta-silicene
or penta-germanene) is shown in Figure~\ref{fig1}. The
top views of penta-graphene and penta-silicene (penta-germanene) are
identical and the side views have slightly
difference (distortion)\cite{zhang2015penta}.
There are 6 atoms in the square primitive cell indicated by
the purple shading. Two hybridized types of the chemical bond can be
found in this pentagonal system. One is \textit{sp$^3$} with four
coordination numbers and the other is \textit{sp$^2$} with three
coordination numbers. From another perspective, every 4 pentagons
can form a larger 10-sided shape of ``boat'' in the top view and these
pentagons in the side view like a ``ladder''. The lattice constants are
$\left| \vec{a_1} \right|$ = $\left| \vec{a_2} \right|$ = 3.64~\AA~
for penta-graphene\cite{zhang2015penta},
$\left| \vec{a_1} \right|$ = $\left| \vec{a_2} \right|$ = 5.58~\AA~
for penta-silicene,
and
$\left| \vec{a_1} \right|$ = $\left| \vec{a_2} \right|$ = 5.67~\AA~
for penta-germanene. According to group theory, regular pentagon,
different from the equilateral triangle, quadrangle, and hexagon, can
not be periodically arranged in a whole plane. Hence, all 2D pentagon
has an intrinsic buckling to keep energetic
stability\cite{ressouche2009magnetic}. As the atomic radius increases in
group IV element, the buckling distance \textit{d} reasonably increases
from 1.2~\AA~for penta-graphene\cite{zhang2015penta} to 2.44 ~\AA~and
3.32~\AA~for penta-silicene and penta-germanene, respectively.

In order to verify the stabilities of penta-silicene and penta-germanene,
we calculated the phonon dispersions shown in Figure~\ref{fig2}(a) and
Figure~\ref{fig2}(b). There are 18 phonon branches due to 6 atoms
in the primitive cell. Owing to the membrane effect, 2D materials have
a linear transverse acoustic (TA), a linear longitudinal acoustic (LA),
and a quadratic out-of-plane acoustic (ZA) phonon modes around the
$\Gamma$ point\cite{gao2018unusually}. All phonon frequencies are free
from the negative values, indicating dynamical stabilities. Since the
averaged phonon frequency is inversely proportional to the atomic mass,
the maximum vibrational frequencies are significantly suppressed from
the penta-graphene\cite{zhang2015penta} of 1666.5~cm$^{-1}$ to the
penta-silicene of 497.7~cm$^{-1}$ and finally to the penta-germanene
of 250.7~cm$^{-1}$, indicating a gradually decreasing trend of the
harmonic phonon strength and interatomic bonding.

A large frequency gap between acoustic and optical phonons (\textit{a-o}
gap) generally leads to a high lattice thermal
conductivity $\kappa$\cite{li2018high}, such as cubic boron
arsenide with
2240 W/mK (\textit{a-o} gap $\approx$ 306.6~cm$^{-1}$)\cite{lindsay2013first}
and 
MoS$_2$ with
103 W/mK (\textit{a-o} gap $\approx$ 52~cm$^{-1}$)\cite{cai2014lattice}.
From the opposite side, a small \textit{a-o} gap is highly desirable for
designing ultralow $\kappa$ materials. The zero \textit{a-o} gap is
found in Figure~\ref{fig2}(a) and Figure~\ref{fig2}(b) for penta-silicene
and penta-germanene. This imply strong acoustic-optical phonon
scattering and ultralow $\kappa$ in penta-silicene and
penta-germanene\cite{gao2018unusually}.

Group velocity, defined by the $\vec{\upsilon}=d \omega/d \vec{q}$,
is one of the key parameters in determining the final $\kappa$.
Because it is a vector (negative sign represents the opposite
direction), we use the scalar $\left| \vec{\upsilon} \right|^2$
to study the phonon transport of penta-silicene and penta-germanene
with mode resolution along x-axis and y-axis, shown in
Figure~\ref{fig2}(c) and Figure~\ref{fig2}(d). Penta-silicene has
a larger $\left| \vec{\upsilon} \right|^2$ than penta-germanene
no matter along the x-axis and y-axis. Along both of them, larger
$\left| \vec{\upsilon} \right|^2$ are found in acoustic phonon
modes (colorful dot/cubic) than optical phonon modes (black dot/cubic),
indicating a more dispersive behavior of acoustic branches compared
to the 
optical phonon modes. The LA mode of penta-silicene
along $\Gamma$-X is quite different from of it along $\Gamma$-Y shown
in Figure~\ref{fig2}(a) and Figure~\ref{fig2}(c), suggesting a large
anisotropic phonon property along x-axis and y-axis. However,
penta-germanene is almost isotropic shown in Figure~\ref{fig2}(b)
and Figure~\ref{fig2}(d).
%
On the whole, the phonon transport at the group velocity level is
significantly suppressed for penta-silicene and penta-germanene
compared with the penta-graphene\cite{wang2016lattice}.

\begin{figure}[t!]
\includegraphics[width=1.0\columnwidth]{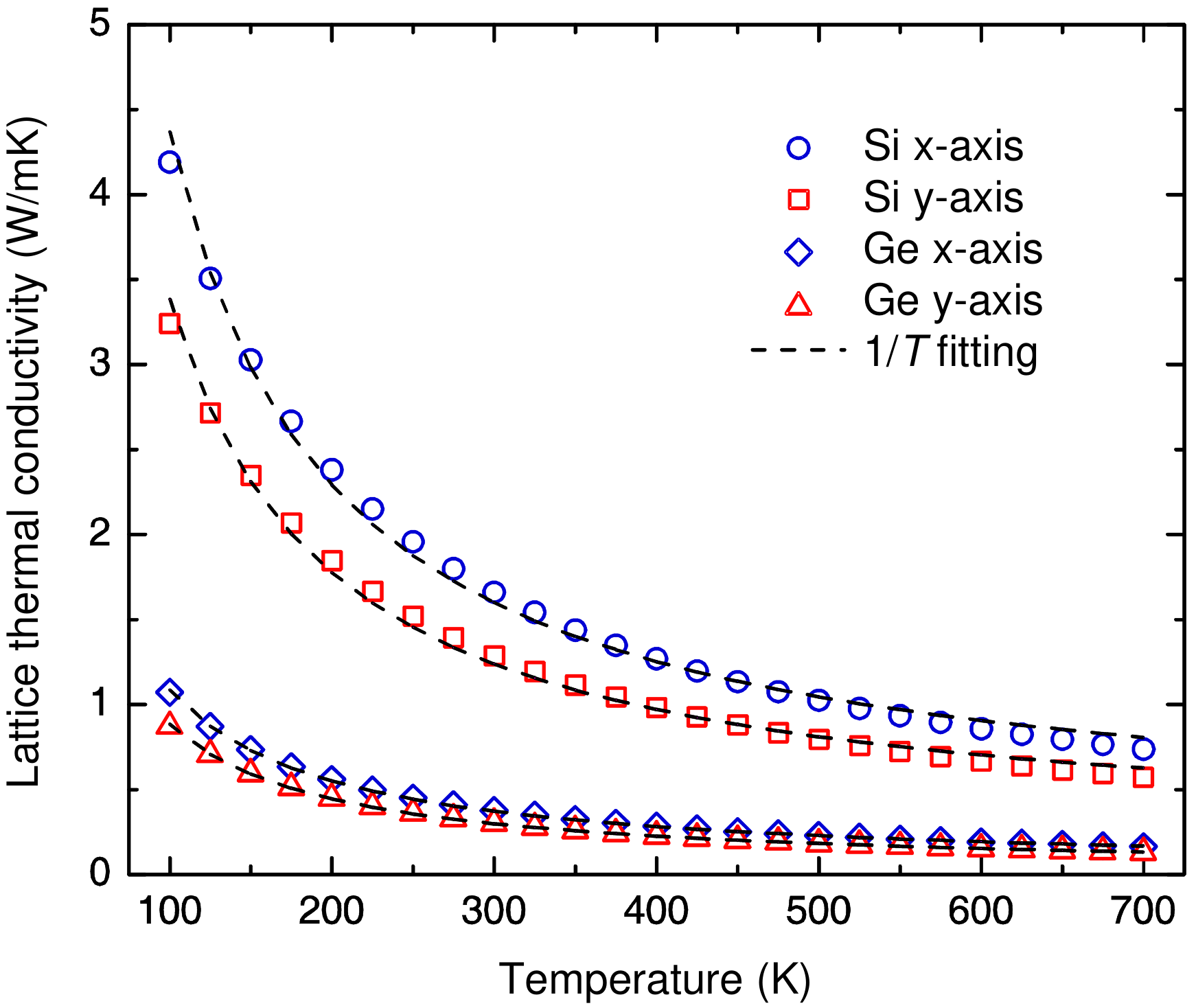}
\caption{Lattice thermal conductivity of penta-silicene and
penta-germanene as a function of temperature along x-axis and
y-axis from the Boltzmann transport equation. 1/\textit{T}
fittings are shown in the dashed lines, indicating a dominant
Umklapp process of phonon-phonon scattering that brings about
the thermal resistivity.
}\label{fig3}
\end{figure}

In undoped semiconductors and insulators, phonons,
rather than electrons, are the main carrier for the heat transport.
In the framework of relaxation time approximation (RTA) and Boltzmann
equation, lattice thermal conductivity $\kappa$ with phonon modes
$\lambda$ and wave vector \textbf{q} can be obtained\cite{gao2018unusually}:
\begin{equation} %
\label{kappa}
\kappa_{\alpha\beta} = \frac{1} {V} \sum_{\lambda} C_\lambda
                       \upsilon_{\lambda \alpha}
                       \upsilon_{\lambda \beta} \tau_{\lambda},  \\
\end{equation}
in which \textit{V} is the volume of the primitive cell,
\textit{C$_\lambda$}, $\tau_\lambda$, and $\upsilon_{\lambda \alpha}$
are the specific heat, relaxation time, and group velocity
in the Cartesian direction $\alpha$ of each single phonon
mode $\lambda$ ($\nu$, \textbf{q}), respectively. Generally,
there are two types of phonon-phonon scattering. One is the
Umklapp (U) process and the other is the Normal (N) process.
The former is the only contributor to thermal resistance.
To solve the single-mode RTA equation, the N process can not be
excluded from the whole scattering rates and is wrongly regarded
as same as the U process. Hence, we use the corrected result
after adopting iterative procedure (removal of the N
process)\cite{li2014shengbte,lindsay2013ab}.

Figure~\ref{fig3} shows the calculated $\kappa$ of
penta-silicene and penta-germanene along the x-axis and y-axis.
Although the lattice constants along the x-axis and y-axis are
the same, the $\kappa$ values of heat transport are obviously
anisotropic. At room temperature, $\kappa$ of penta-silicene
are 1.66~W/mK along the x-axis and 1.29~W/mK along the y-axis.
For penta-germanene, the $\kappa$ are reduced to 0.38~W/mK and
0.30~W/mK along both directions, implying a strong anisotropic
heat transport property. This is quite different
from that of penta-graphene which has an isotropic $\kappa$ of
645~W/mK at 300~K\cite{wang2016lattice}. The anisotropy
ratio ($\kappa_x$/$\kappa_y$) is 1.29 for penta-silicene and
1.25 for penta-germanene, respectively. This phenomenon can be
derived from the anisotropic phonon dispersions and phonon group
velocities shown in Figure~\ref{fig2}. This in-plane anisotropic
property may attract more attention to orientation-dependent thermal
devices\cite{du2016auxetic} and increase the diversity of the
2D pentagonal system in group IV elements.

\begin{table}
 \begin{threeparttable}
		\caption{Relevant thermal properties of silicon and germanium
                 based 2D materials, as well as penta-graphene.
                 $\omega_D^a$ (THz), $\omega_\Gamma^o$ (THz) and
                 $\kappa$ (W/mK) are the largest acoustic phonon
                 frequency (Debye frequency), lowest optical phonon
                 frequency at $\Gamma$ point and lattice thermal
                 conductivity at 300~K, respectively. All intrinsic
                 $\kappa$ is the smaller one if the material is
                 anisotropic along x and y directions.}\label{Tab1}
		\renewcommand\arraystretch{1.5}
		\begin{tabular*}{0.50\textwidth}{p{2.3cm}p{1.5cm}p{1.7cm}p{1.7cm}}		
			\hline \hline
	        Materials           & ~$\omega_D^a$ & ~~~~$\omega_\Gamma^o$ & ~~~~$\kappa$    \\
			\hline
            Silicene$^a$        & 6.00   & ~~~5.70   & ~~9.40           \\
            Germanene$^b$       & 2.91   & ~~~5.10   & ~~2.38           \\
            SiTe$_2$$^c$        & 2.77   & ~~~2.34   & ~~2.27            \\
            Penta-C$^d$         & 14.8   & ~~~17.5   & ~~645            \\
            Penta-Si$^e$        & 3.01   & ~~~2.89   & ~~1.29            \\
            Penta-Ge$^e$        & 1.75   & ~~~1.52   & ~~0.30            \\
			\hline \hline
		\end{tabular*}
 \begin{tablenotes}
        \footnotesize
         \item[a]Hexagonal silicene\cite{xie2014thermal}. $^b$Hexagonal germanene\cite{peng2016phonon}.
         $^c$1T-SiTe$_2$\cite{wang2019ultralow}. $^d$Penta-graphene\cite{zhang2015penta}.
         $^{e}$Present work.
      \end{tablenotes}
    \end{threeparttable}
\end{table}

According to the Slack model for bulk materials, there are two
vital parameters correlated with $\kappa$. One is the largest
acoustic phonon frequency (Debye frequency) $\omega_D^a$, and
the other one is the Gr\"{u}neisen parameter
$\gamma$\cite{slack1973nonmetallic}. Long-wave (acoustic) phonons
are the main contributors to the heat transport. A low $\omega_D^a$,
usually, means a small $\kappa$. 
The calculated $\omega_D^a$ are shown in Table ~\ref{Tab1}. In order
to compare, we also list other silicon and germanium based 2D
materials. Silicene and
germanene are the representatives of hexagonal lattice and
SiTe$_2$ belongs to TMDs\cite{gao2019degenerately} and is regarded
as a superior thermoelectric
material\cite{wang2019ultralow,tshitoyan2019unsupervised}.
Based on Table ~\ref{Tab1}, we can find that $\kappa$ is
proportional to the $\omega_D^a$, verifying the
correctness of the Slack model prediction in 2D materials
qualitatively.

More importantly, penta-silicene and penta-germanene have the
lowest $\kappa$ in all silicon and germanium based 2D materials.
Based on our previous collection\cite{gao2018unusually},
penta-germanene has the smallest $\kappa$ of 0.30~W/mK in the
2D family. This ultralow $\kappa$ is highly desirable for the
ultrahigh thermoelectric material since $\kappa$ is inversely
proportional to the efficiency of conversion from the heat
energy to the electrical energy\cite{gao2018high}. Besides, as
temperature increases, $\kappa$ decreases significantly for the
pentagonal system. The $\kappa$ can be well described by
$\kappa$ $\propto$ 1/\textit{T}, indicating the dominant
phonon-phonon Umklapp scattering in penta-silicene and
penta-germanene. As temperature increases to 600~K, $\kappa$
is further suppressed to 0.86~W/mK and 0.66~W/mK along the
x-axis and y-axis for penta-silicene, while they are
0.19~W/mK and 0.15~W/mK for penta-germanene.

\begin{figure*}[t!]
\centering
\includegraphics[width=1.8\columnwidth]{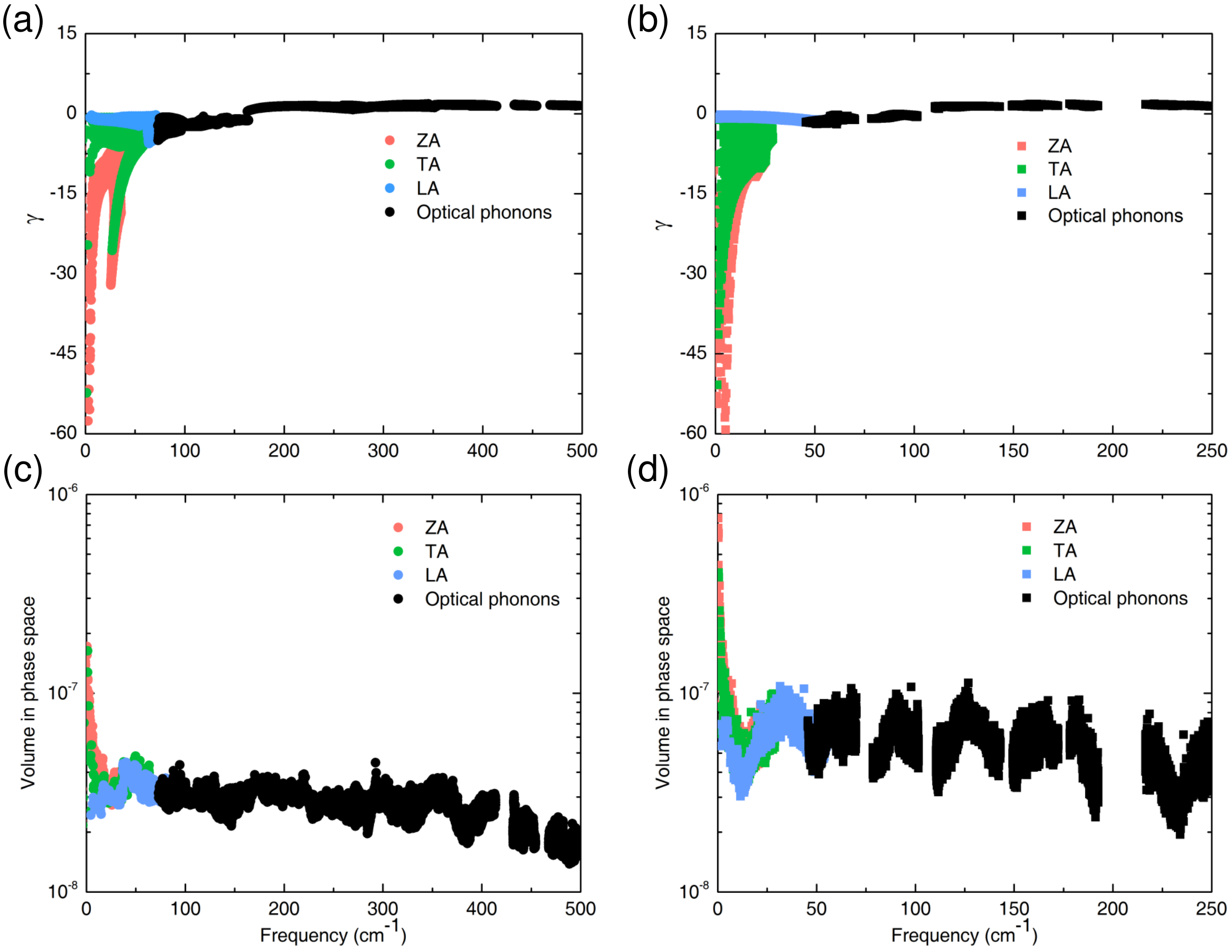}
\caption{The Gr\"{u}neisen parameter $\gamma$ of (a) penta-silicene
and (b) penta-germanene as a function of the phonon frequency with
mode resolution. Phase-space volume \textit{P$_3$} for three-phonon
anharmonic scattering process of (c) penta-silicene and (d)
penta-germanene. \textit{P$_3$} is inversely proportional to the
lattice thermal conductivity $\kappa$ qualitatively.
}\label{fig4}
\end{figure*}

\begin{figure*}[t!]
\centering
\includegraphics[width=1.8\columnwidth]{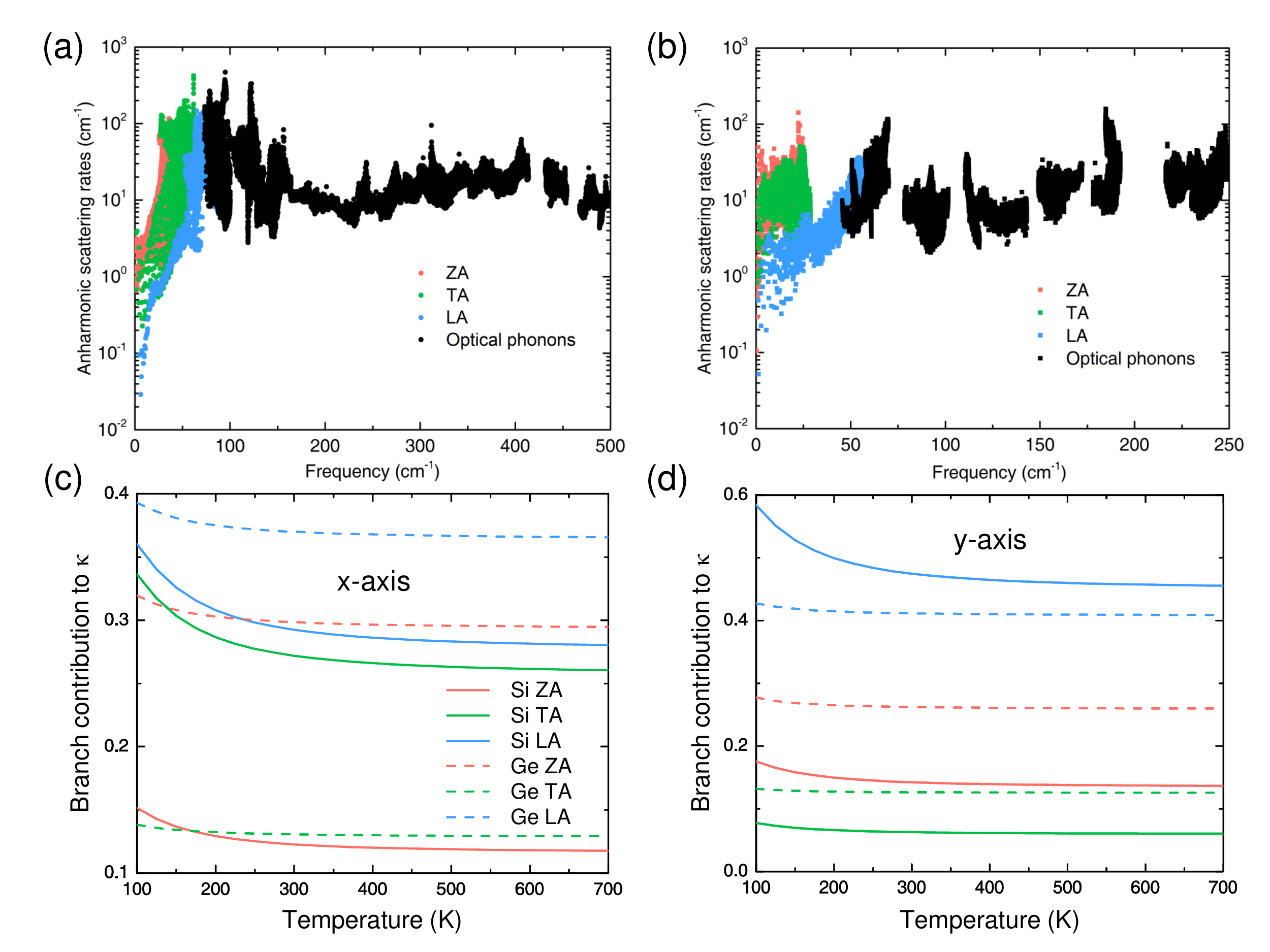}
\caption{Calculated anharmonic scattering rates of (a) penta-silicene
and (b) penta-germanene with mode resolution at 300 K. The normalized
contribution of three acoustic phonon branches to the lattice thermal
conductivity $\kappa$ along the (c) x-axis and (d) y-axis as a
function temperature.
}\label{fig5}
\end{figure*}

According to the Eq. \eqref{kappa} and the Slack model, anharmonic
phonon property also has a large effect on thermal transport and
$\kappa$. Specifically, the phonon relaxation time relies on two
components: (i) the intensity of each phonon-phonon scattering mode.
This quantity can be reflected by the mode-dependent Gr\"{u}neisen
parameter $\gamma$ which is defined as\cite{wu2016hydrogenation}
\begin{equation} %
\label{gruneisen}
\gamma = - \frac{d \textrm{ln} \omega} {d \textrm{ln} V},  \\
\end{equation}
where phonon frequency $\omega$ is a function of band indexes $\nu$,
and wave vector \textbf{q}. (ii) the total number of phonon-phonon
scattering modes. Each available phonon scattering mode must
simultaneously satisfy the energy and quasi-momentum conservations.
This process can be quantitatively described by the volume of the
scattering phase space \textit{P$_3$} for three-phonon
processes\cite{lindsay2008three,lee2014resonant}. These two parameters
finally enter into the three-phonon scattering rates, written as
\begin{equation} %
\label{rates1}
\Gamma_{\lambda \lambda' \lambda''} ^+ = \frac{\hbar \pi}{4} \frac{f'_0-f''_0}{\omega_\lambda \omega_{\lambda'} \omega_{\lambda''}} \left| V^+_{\lambda \lambda' \lambda''} \right|^2 \delta( \omega_\lambda + \omega_{\lambda'} - \omega_{\lambda''} ), \\
\end{equation}
\begin{equation} %
\label{rates2}
\Gamma_{\lambda \lambda' \lambda''} ^- = \frac{\hbar \pi}{4} \frac{f'_0+f''_0 +1}{\omega_\lambda \omega_{\lambda'} \omega_{\lambda''}} \left| V^-_{\lambda \lambda' \lambda''} \right|^2 \delta( \omega_\lambda - \omega_{\lambda'} - \omega_{\lambda''} ), \\
\end{equation}
where $\Gamma_{\lambda \lambda' \lambda''}^+$ and $\Gamma_{\lambda \lambda' \lambda''}^-$
stand for the absorption and emission processes\cite{li2014shengbte}.
The Gr\"{u}neisen parameter $\gamma$ is proportional to the scattering
matrix elements $\left| V^{\pm}_{\lambda \lambda' \lambda''} \right|$
and the number of Dirac delta distributions is equal to \textit{P$_3$}.
We can find that three-phonon scattering rates
$\Gamma_{\lambda \lambda' \lambda''}$ are closely positive correlated with
$\gamma$ and \textit{P$_3$}. We will discuss the anharmonic phonon behavior
of the 2D pentagonal system from the above two important factors.

According to the Eq. \eqref{gruneisen}, $\gamma$ provides the
anharmonic phonon property and a large $\gamma$ means a large
anharmonicity. The calculated $\gamma$ of penta-silicene
and penta-germanene with mode resolution is shown in
Figure~\ref{fig4}. It is found that acoustic phonon modes for
both materials have large negative $\gamma$, while optical
phonon branches have small positive $\gamma$. Like graphene,
a negative sign of $\gamma$, generally implies a negative
thermal expansion that will mitigate the thermal strain and
stress in high-temperature electronic
devices\cite{liu2019anisotropic,liu2019strain}. Based on the previous work,
$\gamma$ of ZA mode for penta-graphene is large, however,
$\gamma$ of TA and LA phonon modes are quite small (around
zero)\cite{wang2016lattice}. This situation is
different from that of penta-silicene and penta-germanene.
$\gamma$ of TA mode in penta-silicene (green dot) has the
averaged value of around $-10$ but smaller than that of
penta-germanene with an averaged value of around
$-15$ (green cubic). From this angle, penta-germanene has
the largest anharmonic interactions with mode resolution
in the 2D pentagonal system.

The calculated \textit{P$_3$} of penta-silicene and
penta-germanene are shown in Figure~\ref{fig4}(c) and
Figure~\ref{fig4}(d). Evidently, \textit{P$_3$} of
penta-germanene is larger than that of penta-silicene,
indicating a larger number of phonon-phonon scattering
channels. \textit{P$_3$} of penta-silicene is smaller
than that of penta-graphene\cite{wang2016lattice}.
Therefore, we can conclude that the difference of
$\kappa$ between penta-silicene and penta-graphene
originates from the Gr\"{u}neisen parameter $\gamma$
and group velocity $\upsilon$, rather than the volume
in phase space \textit{P$_3$}. Specifically, a larger
$\gamma$, a larger \textit{P$_3$}, and a smaller
$\upsilon$ lead to a smaller $\kappa$ of penta-germanene
compared to the penta-silicene.

The above two independent factors ($\gamma$ and
\textit{P$_3$}) finally lead to one vital parameter called
three-phonon scattering rates (reciprocal of the relaxation time)
shown in Eq. \eqref{rates1} and Eq. \eqref{rates2}.
This quantity directly enters the $\kappa$ shown in
Eq. \eqref{kappa}. A large anharmonic scattering rate needs
a large $\gamma$ and \textit{P$_3$} concurrently.
The calculated anharmonic scattering rates of
penta-silicene and penta-germanene are shown in
Figure~\ref{fig5}(a) and Figure~\ref{fig5}(b). Overall,
the anharmonic scattering rates of both are
comparable. However, the anharmonic scattering rates of
LA mode in Figure~\ref{fig5}(a) in penta-silicene are
quite larger than that of penta-germanene
Figure~\ref{fig5}(b). There is an interesting region of
large acoustic-optical phonon scattering located from
40~cm$^{-1}$ to 90~cm$^{-1}$ in penta-silicene,
indicating strong phonon-phonon interactions between
acoustic phonons and optical phonons. This originates
the special phonon dispersion shown in
Figure~\ref{fig2}(a). Compared with the penta-germanene,
penta-silicene has a larger cross-region between acoustic
phonons and optical phonons, which makes more easier to
simultaneously satisfy the energy and quasi-momentum
conservations.

\begin{figure*}[t!]
\centering
\includegraphics[width=1.8\columnwidth]{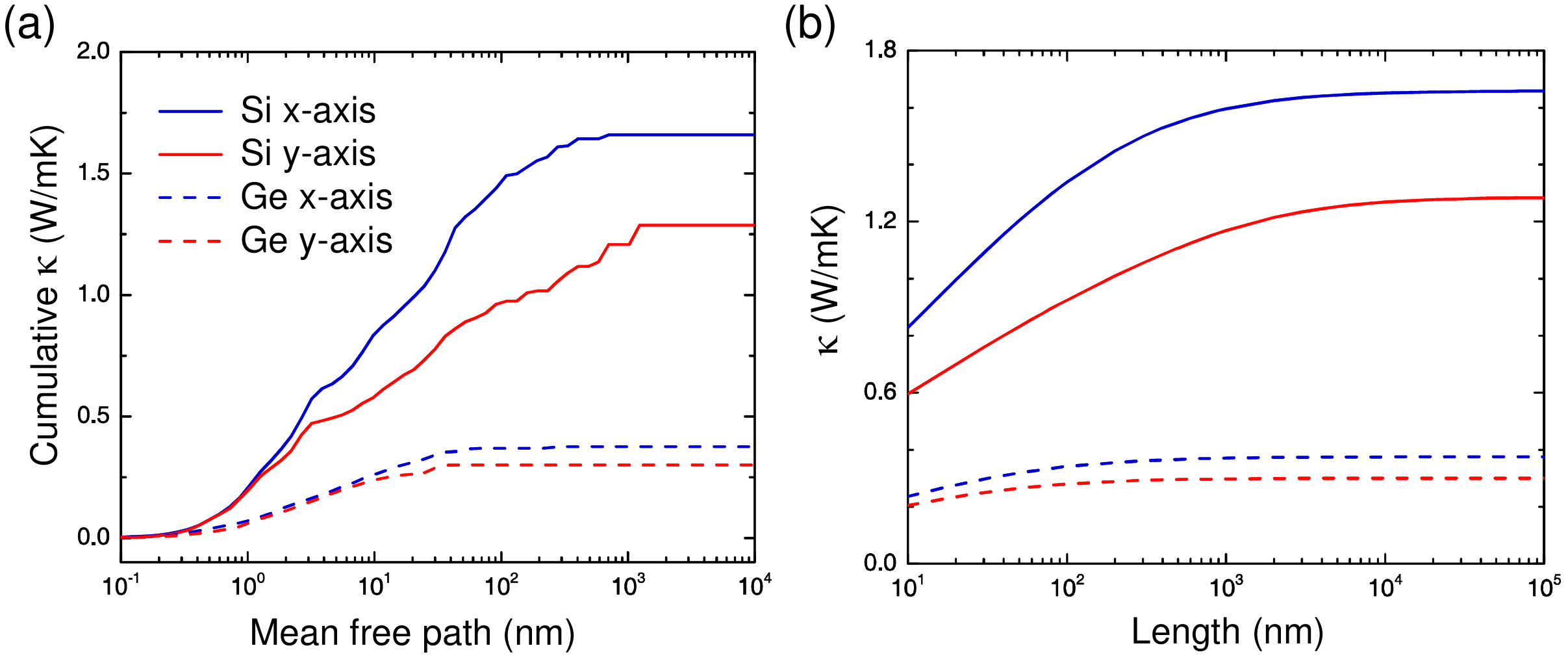}
\caption{(a) The cumulative lattice thermal conductivity $\kappa$
as a function of the phonon mean free path of 
penta-silicene (solid lines) and penta-germanene (dashed lines)
along x-axis and y-axis at 300 K. (b) $\kappa$ as a function of
sample size based on the Eq. \eqref{boundary} at 300 K.
}\label{fig6}
\end{figure*}

To further explore the underlying phonon scattering mechanism,
we plot the mode contribution of three acoustic phonons to
the $\kappa$ for penta-silicene and penta-germanene shown
in Figure~\ref{fig5}(c) and Figure~\ref{fig5}(d). Only two
types of scattering processes are allowed here. One is
the scattering within three acoustic phonons and the other
one is the scattering between two acoustic phonons (one acoustic
phonon) and one optical phonon (two optical phonons). In x-axis,
LA and TA modes dominate the heat transport and ZA mode only
contributes around 12.5\% in penta-silicene at 300~K. For penta-germanene,
LA mode contributes 38\% and ZA mode contributes 30\% to the
$\kappa$ at the same temperature. This is quite different from that
in graphene where 75\% $\kappa$ originates from the ZA phonon
mode\cite{lindsay2010flexural}.
In the y-axis, the situation is a little different. LA mode still
contributes most to the $\kappa$ in both materials while the TA's
contribution is quite small. This anisotropic property of
phonon mode contribution can be traced back to the anisotropic
phonon group velocity shown in Figure~\ref{fig2}.

Furthermore, we also evaluate the cumulative $\kappa$ with
respect to the phonon mean free path for both penta-silicene
and penta-germanene along x and y-axis at 300~K shown in
Figure~\ref{fig6}(a). In order to obtain the characteristic
length \textit{l$_0$}, we use a single parametric
function\cite{li2014shengbte}
\begin{equation} %
\label{cumulative}
\kappa(l \leq l_{max}) = \frac{\kappa_0} { 1+ l_0/l_{max} },  \\
\end{equation}
in which \textit{l$_0$} is the only parameter to be determined.
\textit{l$_{max}$} and $\kappa_0$ are the maximum mean free path
and ultimately cumulative $\kappa$. The calculated cumulative
$\kappa$ of penta-silicene and penta-germanene are shown in
Figure~\ref{fig6}. The fitted parameters, for penta-silicene
are 9.95~nm and 13.69~nm along the x-axis and y-axis. Similarly,
\textit{l$_0$} 
are 4.18~nm and 3.23~nm for penta-germanene along both
directions. This characteristic length \textit{l$_0$} can be
regarded as a representative mean free path of materials. A
low \textit{l$_0$} means a low mean free path of phonon-phonon
scattering. Due to Moore's Law, the size of the electronic
devices is still decreasing and the role of heat transport in
these nanomaterials is becoming increasingly more crucial
to modern transistors. Besides, \textit{l$_0$} can be
evaluated the different types of phonon behavior such as
ballistic transport, diffusive transport, super-diffusive
transport, and hydrodynamics\cite{lee2015hydrodynamic,
cepellotti2015phonon,gao2016heat,gao2016stretch}. In
penta-graphene, \textit{l$_0$} equals 278~nm when
cumulative $\kappa$ reaches 50\% of the total $\kappa$\cite{wang2016lattice},
while
the values of \textit{l$_0$} reduced significantly to
14.2~nm for penta-silicene and 3.2~nm for penta-germanene
along the y-axis, respectively. Furthermore, the $\kappa$ of
penta-silicene and penta-germanene can be further suppressed
by inducing the phonon-boundary scattering and isotope effect
scattering.

In reality, electronic devices must have a length scale.
Hence, 
scattering between boundary and
phonons becomes a very important factor in thermal transport
at the nanoscale. Specifically, the boundary scattering has
been well described and verified in 2D materials, such as in
graphene\cite{balandin2008superior}. The semi-empirical formula
can be written as\cite{balandin2008superior,nika2009phonon}
\begin{equation} %
\label{boundary}
\frac{1}{\tau_b} = \frac{\upsilon_{\nu q}}{L},  \\
\end{equation}
in which \textit{L} and $\nu_{\nu q}$ represent the material
size and phonon group velocity. This part of the scattering
rate will be added to the phonon-phonon scattering rate using
the Matthiessen's rule
$1/\tau_{total}=1/\tau_{p-p}+1/\tau_{b}$\cite{kang2017ionic}.
The calculated $\kappa$ of penta-silicene and penta-germanene
as a function of sample length \textit{L} are shown in
Figure~\ref{fig6}(b). As \textit{L} decreases from 100~nm to
10~nm, the $\kappa$ reduces following an exponential function
$\kappa \propto $ \textrm{log} \textit{L} due to the strong
boundary effect, which has been experimentally verified in
suspended graphene in 2014\cite{xu2014length}.

For an instance,
when \textit{L} is equal to 1~$\mu$m, the $\kappa$ of
penta-silicene are reduced to 1.59~W/mK and 1.17~W/mK along the
x and y-axis, while the $\kappa$ of penta-germanene are suppressed
to 0.37~W/mK and 0.29~W/mK along both directions. When \textit{L}
equals 100~nm, the $\kappa$ of penta-silicene has 80\% and 72\%
percents along the x and y-axis compared with the $\kappa$ of
material with infinite size. Similarly, the $\kappa$ of
penta-germanene has 90\% and 93\% percents along x and
y-axis. In this sense, the $\kappa$ of penta-silicene is more
sensitive to the sample length \textit{L} with respect to the
penta-germanene. The reason is that, for an instance, the characteristic
lengths \textit{l$_0$} of penta-silicene and penta-germanene
are 14.2~nm and 3.2~nm along the y-axis. Hence, at the same
length (for example, 1$\mu$m), more phonons can be scattering
by the boundary in penta-silicene than that of penta-germanene.

We also calculate the electronic thermal conductivity
based on the Wiedemann–Franz law, we find that electronic thermal
conductivity is much lower than lattice thermal conductivity and
can be neglected for low concentration doping (n<10$^{12}$ cm$^{-2}$).
As the carrier concentration doping increases, the electronic
conductivity increases and electronic thermal conductivity
increases simultaneously (10$^{12}$ cm$^{-2}$<n<10$^{13}$ cm$^{-2}$)
due to the population enhancement. For a much higher concentration
doping (n>10$^{13}$ cm$^{-2}$), both materials behavior like a
good ``metal'' having free carriers. Hence, electronic thermal
conductivity increases significantly as a function of carrier
concentration.

\section{Conclusion}
In summary, based on the \textit{ab~initio} calculations and the
Boltzmann equation, we have, firstly, obtained the lattice thermal
conductivities and thermal transport properties of penta-silicene
and penta-germanene, belonging to the 2D pentagonal system in group
IV element. We have found that the lattice thermal conductivities
of penta-silicene and penta-germanene are 1.29~W/mK and
0.30~W/mK, respectively, which are much smaller than that of
penta-graphene of 645~W/mK at room temperature\cite{wang2016lattice}.
More importantly, penta-germanene, to the best of our knowledge, may
have the lowest lattice thermal conductivity in 2D crystal
materials by far\cite{gao2018unusually}. It could be favorable if
this prediction can be further confirmed by other independent
methods, such as molecular dynamics. This ultralow lattice thermal
conductivity can be traced to low phonon harmonic interaction and
strong anharmonic scattering. A small phonon group velocity and a
small Debye frequency indicate a weak phonon harmonic interaction.
A large Gr\"{u}neisen parameter implies a strong phonon scattering
per each phonon mode and a big volume in phase space means a large
number modes available for phonon-phonon interplay. Together these
parameters finally lead to the ultralow lattice thermal conductivity
of penta-silicene and penta-germanene in the 2D pentagonal system.

Although penta-silicene and penta-germanene are metastable structures,
penta-silicene nanoribbon has been successfully synthesized in 2016 and
has shown exotic phenomena, such as topologically protected phases or
increased spin--orbit effects\cite{cerda2016unveiling,prevot2016si}.
Besides, previous work also theoretically verified that penta-silicene
can be stably grown on the Ag surface\cite{guo2019lattice}, which means
a proper confining material (e.g., a metal substrate) is needed to produce
the titled penta-silicene and penta-germanene in experiments.
In this sense, we hope more experimentalists could make attempts on the 2D
pentagonal system such as by using the reactive molecular beam epitaxy
method\cite{ji2019freestanding}. The ultralow thermal conductivity
of penta-silicene and penta-germanene may make a contribution to the
thermal transport of 2D materials and silicon's energy harvesting
power\cite{hu2019silicon}.

\section{Computational method}
The equilibrium geometry and structural stability were
calculated by density functional theory (DFT) implemented
in the VASP code\cite{kresse1996efficient,kresse1996efficiency}.
The exchange-correlation functional of
Perdew-Burke-Ernzerhof (PBE)\cite{perdew1996generalized} was
used. A plane-wave cutoff energy of 500 eV was adopted and a
Monkhorst-Pack Brillouin zone was sampled by 11 $\times$ 11.
The total energy threshold and Hellmann-Feynman forces
between adjacent optimization were 10$^{-6}$ eV and
10$^{-3}$ eV/ \AA. To eliminate spurious interactions
between periodic slabs, a vacuum separation distance of
20~\AA~was applied. Phonon frequency was calculated in
Phonopy\cite{togo2008first} with a 5 $\times$ 5 supercell.
Anharmonic interatomic force constants (IFCs) were extracted
in ShengBTE\cite{li2014shengbte} by solving the linearized
Boltzmann transport equation. The converged $\kappa$ was
obtained after careful parameter testing. The interaction
cutoff was 0.55~nm and the $\Gamma$-centered q-grid was
101 $\times$ 101. A scale broadening parameter of 0.1 for Gaussian
smearing was used. Since thickness for 2D material is not
well-defined, an effective thickness should be chosen to compare
with the 3D material. Here, the effective thickness \textit{h}
is defined as the summation of buckling distance \textit{d} and two
van der Waals radii of the outmost surface atom of
structure\cite{gao2017novel,gao2018two}. For penta-silicene and
penta-germanene, \textit{d} are 2.44~\AA~and 3.32~\AA, while
\textit{h} are 6.64~\AA~and 7.54~\AA~, respectively.

\section*{Conflicts of interest}
There are no conflicts to declare.

\quad\\
{\noindent\bf Author Information}


{\noindent\bf ORCID}\\
Zhibin Gao: 0000-0002-6843-381X \\
Zhaofu Zhang: 0000-0002-1406-1256

\section*{Acknowledgements}
We thank Jiongzhi Zheng for helpful discussions. This work
is supported by an MOE tier 1 grant R-144-000-402-114.

\providecommand*{\mcitethebibliography}{\thebibliography}
\csname @ifundefined\endcsname{endmcitethebibliography}
{\let\endmcitethebibliography\endthebibliography}{}

\end{document}